\documentstyle{aipproc}
 
\def\pa{\partial}

\def\a{\alpha} 
\def\b{\beta} 
\def\d{\delta} 
\def\e{\epsilon}

\def\m{\mu} 
\def\n{\nu}

\def\s{\sigma}

 \def\O{\Omega}
\def\be{\begin{equation}}
\def\ee{\end{equation}}
     
\begin{document}

\title{Black Hole Electromagnetic Duality}

\author{S. Deser}
\address{Department of Physics, Brandeis University,
Waltham, MA 02254, USA}
\maketitle
\begin{abstract}
After defining the concept of duality in the context
of general $n$-form abelian gauge fields in 2$n$
dimensions, we show by explicit example the 
difference between apparent but unrealizable 
duality transformations, namely those in $D=4k+2$,
and those, in $D=4k$, that can be
implemented by explicit dynamical generators.  
We then consider duality transformations
in Maxwell theory in the presence of gravitation, 
particularly electrically and magnetically 
charged black hole geometries.  By comparing
actions in which both the dynamical variables and
the charge parameters are ``rotated," we show their
equality for equally charged electric and magnetic 
black holes, thus establishing
their equivalence for semiclassical processes which
depend on the value of the action itself.
\end{abstract}

I begin this lecture by paying my respects to the 
memory of Juan Jos\'{e} Giambiagi, to whom this
conference is dedicated.  Having known him since the
early '60's, I have had the opportunity of 
understanding his importance not only through his 
physics (universal as many of his ideas have become)
but also through the inspiration he provided in the
evolution of physics research in Argentina and indeed
throughout all Latin America.  He was a man of great
culture, with both knowledge and perspective across
a wide spectrum of human ideas, and a man of great
courage as I was able to observe in the dark days
around 1970 when he was exiled to La Plata.  He 
was an optimist in spite of his dark insights. We 
will all miss him.

An earlier important loss to Latin American 
physics was that of Carlos Aragone of Uruguay and
Venezuela, with whom I had the pleasure of a 25
year collaboration.  He was another leader of 
theoretical physics in our far-flung community,
who twice helped create fruitful environments --
in his original and in his adopted homelands.

Finally, I thank the organizers for inviting me,
even though my topic is not in the mainstream
of this conference.  The work described here was 
performed in collaboration with M. Henneaux and 
C. Teitelboim.  Indeed it builds on work first
done with the latter \cite{001} some 20 years ago!
It will appear in Phys. Rev. D early in 1997 
\cite{002}, and in another paper still in process, 
from which the general $n$-form discussion is
drawn.

Our motivation for returning to so old a topic is 
its relevance to current research.  I will have
time here to discuss only one aspect of duality, 
namely its application to 
black hole physics, particularly that of charged
black holes and their semiclassical behavior, that
is when the actions themselves (I/$\hbar$) and not 
just the field equations matter.  Since both
electrically and magnetically charge black holes
can exist, investigating their equivalence in this
regime is tantamount to establishing a generalized
Maxwell duality in presence of sources, both 
electric and magnetic, as well as of a 
gravitational field.  We will indeed show
(after reviewing the flat space, sourcefree case)
that the actions of magnetically and electrically
(equally) charged black holes are in fact the
same, a conclusion recently reached in \cite{003} 
by very different means.\footnote{This is not directly 
related to the very different 
question of charge quantization in the {\it e.g.},
$\sim \hbar$ sense.  Also, we will be considering
here the fixed charge sectors rather than the
complementary case of fixed chemical potential, but 
the results should carry through to that situation
as well.}

Let me begin with some introductory notions about
duality in a more general framework to show also
what duality is {\it not}, as there are still a
number of misconceptions in the literature.
Consider a general ($n$--1)-form potential and 
its associated field strength $F_{1..n} \equiv
\pa_n \, A_{1..n-1}$.  [All potential and field
indices are to be understood to be totally
antisymmetrized and suitably normalized; 
also I use ``mostly plus" metric
signature.]  The dual of a field is always 
defined to be
\be
^*\!F^{1..n} \equiv \frac{1}{n!} \: 
\e^{1..n \, n+1..2n}
F_{n+1..2n}
\ee
where $\e$ is the Levi--Civita symbol (with 
$\e^{01..} = +1$) in 2$n$ dimensions. Clearly 
only in 2$n$ dimensions will $n$-form fields be
of the same rank as their duals so that one can 
even attempt to speak of duality transformations, 
let alone invariances.  Now the action, field 
equations, and Bianchi identities for a 
source-free field are
\be
I = -c_n \int d^{2n}x \, F_{1..n} F^{1..n}, 
\hspace{.3in}
\pa_1 F^{1..n} = 0 , \hspace{.3in} \pa_1 \, 
^*\!F^{1..n} \equiv 0
\ee
where $c_1 = 1/2$, $c_2 = 1/4$ etc.  The 
(source-free) field equations and Bianchi 
identities are of the same form so that formally
any linear transformation
\be
F \rightarrow aF + b\, ^*\!F
\ee
together with its
dual, $^*\!F \rightarrow a^*\!F + b~^{**}\!F$ also
gives $F$'s that
obey this pair of equations.  Double duality
is an operation that depends on whether 
$n = d/2$ is even or odd, as a little reflection
on the $\e$ symbol verifies:
\be
^{**}\!F = F, \;\; n = 2k+1, \hspace{.4in} 
^{**}\!F = -F, \;\; n = 2k
\ee
(this is also the reason self-duality is only
realizable in the $n = 2k+1$ case).  Either
way, the above formal transformation is compatible
with the equations.  Is this symmetry shared
by other physical quantities of these theories,
in particular by their actions (our main
interest here) and by their stress-tensors?
Although it is only the Poincar\'{e} generators 
that are physical in flat space, the local 
stress tensor becomes an observable current
in presence of gravity.  These quantities are
bilinear in the fields so they should impose more
stringent conditions than the -- linear -- 
equations.  To see most clearly what restrictions
on (3) they impose let us rewrite the bilinears
symmetrically in terms of $F$ and $^*\! F$.
Surprisingly, the actions and stress tensors are
of the same form in all dimensions, because
the scalar identity $F_{\m ..}F^{\m ..} \equiv
-^*\! F_{\m ..} ^*\! F^{\m ..}$  is by (4)
dimension-independent.  It then follows from
(2) that
$$
I = - \frac{1}{2} c_n \int d^{2n}x 
(F^2 - ^*\!\! F^2 )\; . 
\eqno{\rm (5a)} 
$$
The corresponding stress-tensors are then easily
found, by varying with respect to the metric in
the usual way:
$$
T^\m_\n = \frac{1}{2} 
(F^{\m ..} F_{\n ..} + ~ ^{*}\!\! F^{\m ..}
~^{*}\! F_{\n ..}) \; .
\eqno{\rm (5b)} 
$$
In accordance with conformal invariance of the
action, $T^\m_\m = \frac{1}{2} (F^2 +
^*\!F^2 ) \equiv 0$. In all cases, there is the 
same ``mismatch" between the 
signs in the action and stresses, so that not
both would seem to remain invariant under a duality 
transformation.  The latter must be defined as either
a normal rotation or a hyperbolic one 
rather than the general (3) to even
formally keep either a sum or a difference of
squares invariant.  There is also no help from the
fact that cross terms in the form $F^*\! F$
are total divergences and hence irrelevant to the 
action (apart from possible topological effects).
That is, in 4$k$ dimensions 
$F_{\m\n ..}~^*\!F^{\m\n ..} = \pa_\m 
[\e^{\m ..} A \pa A]$ is the divergence of a
Chern--Simons structure, while in 4$k$+2, $F^*\! F$ 
actually vanishes identically, {\it e.g.},
$F_\m (\e^{\m\n}F_\n ) \equiv 0$. So we have a 
paradox: the equations and identities
in all dimensions are together invariant under
any linear variation of $F$ and $^*\! F$ into
each other, while the action and stress tensor
can seemingly never both be invariant under any 
transformation at all.  
In fact, as we will now show, none of the 
above considerations is even meaningful and
(despite the uniformity in (5a) and (5b))
the correct answer is that Maxwell theory
and its 4$k$ extensions are 
perfectly invariant in a precise sense under 
duality rotations, while duality is not 
even definable for scalar theory 
and {\it its} (4$k$+2) generalizations.

The basis for those statements is the simple
remark that in a dynamical theory, only
transformations that can be generated by
functionals of the canonical variables are 
even meaningful. Until the latter are given,
one cannot even know what (if any) duality
change is possible, let alone whether it
defines an invariance.  Thus, the scalar 
field in $D$=2,
\renewcommand{\theequation}{\arabic{equation}}
\setcounter{equation}{5}
\be
I = - \frac{1}{2} \int d^2x \, F_\m F^\m \hspace{.4in}
F_\m \equiv \pa_\m \phi  \hspace{.4in}
^*\!F^\m \equiv \e^{\m\n} \pa_\n \phi 
\ee
has Hamiltonian form 
\be
I = \int d^2x [\pi \dot{\phi} - \frac{1}{2} 
(\pi^2 + \phi^{\prime 2})] \; ,
\ee
the field strength having components
$F_0 = \dot{\phi} = \pi$, $F_1 = \phi^\prime$.
Now it is clear that there is no generator
$G = \int dx {\cal G} (\pi ,\phi )$ such that
its Poisson bracket with $\pi$ and $\phi^\prime$
will rotate them into each other
(with either sign).  For example
$[G, \pi (x)] \sim \phi^\prime (x)$ would require
$G \sim \int dy \, \phi (y) \phi^\prime (y)$
but that is clearly a total divergence and
similarly for $[G, \phi^\prime ] \sim \pi$.  It
is easy to see 
(by counting signs in $\e$) that this impossibility 
extends to the general $D=4k+2$ case.\footnote{Eq. (12)
immediately shows that $\e^{ijklm} A_{ij} \pa_k
A_{lm}$ is a total divergence for even form potentials
represented here by $A_{ij}$.}

Let us turn to  $D=4k$, in 
particular to electrodynamics in $D=4$, our main 
topic. We start with a quick review of the flat 
space source-free sector \cite{001}.  Here the 
Maxwell action may be written in terms of the 
reduced first order conjugate variables 
({\bf E},{\bf A}) as
\be
I_M [{\bf E},{\bf A}] = \int d^4x [-{\bf E}
\cdot \dot{\bf A} - \frac{1}{2}
({\bf E}^2 + {\bf B}^2)]\; ,
\hspace{.3in} \mbox{\boldmath $\nabla$}
\cdot {\bf E} = 0 \; ,
\ee
where ${\bf B} \equiv \nabla \times {\bf A}$.
In the absence of sources, the Gauss constraint
says that {\bf E} is purely transverse,
\be
{\bf E} \equiv \mbox{\boldmath $\nabla$} \times
{\bf Z}
\ee
and therefore only the transverse, gauge-invariant,
part of $\dot{\bf A}$ survives in the kinetic
term, which may be rewritten as
\be
\int d^4x \: \e^{ijk} \pa_j Z_k \dot{A}_i \; .
\ee
We assert, and it is easy to check, that the 
above reduced $I_M$ is invariant under the 
rotation of the 2 dimensional vector with 
components $V \equiv ({\bf Z},{\bf A})$ or its
curl $W \equiv ({\bf E},{\bf B})$ under
the usual 2-dimensional rotation,
\be
V^\prime = RV \hspace{.4in} \mbox{or}
\hspace{.4in} W^\prime = RW \; , \hspace{.4in}
R \equiv \exp (i \s_2 \cos \theta ) \; .
\ee
Equally important is that the generator of this
transformation exists and has a very elegant
``topological" (metric independent) Chern--Simons
form,
\be
G = - \frac{1}{2} \int d^3x \: \e^{ijk} 
[Z_i \pa_j Z_k + A_i \pa_j A_k] \; .
\ee
The Poisson bracket or commutator of $G$ with $V$ or 
with $W$ engenders (11) by virtue of the canonical
commutation relations $[{\bf E}^i,{\bf A}^\prime_j]
= [\d^i_j ({\bf r}-{\bf r}^\prime)]^T$ where $\d^T$ 
is the usual
transverse projection of the unit operator.  As
usual there is some asymptotic falloff to be 
specified; here and in curved space we take
${\bf A} \sim a(\O ) r^{-1} + {\cal O} (r^{-2})$
and ${\bf E} \sim {\bf e} (\O ) r^{-2} + 
{\cal O} (r^{-3})$ where ${\bf a},{\bf e}$ depend
only on solid angle.

We must now generalize the above analysis to include
nontrivial geometries and charges.  The former is
easy: Just write the Maxwell action in the covariant
first order form,\footnote{This can be done in the 
same way for all form actions and incidentally 
exhibits their common Weyl invariance.}
\be
I_M = - \frac{1}{2} \int d^4x 
\left[ F^{\m\n} (\pa_\m A_\n - \pa_\n A_\m )
- \frac{1}{2} \: F^{\m\n} 
F^{\a\b} g_{\m\a}g_{\n\b} (-g)^{-1/2} \right]
\ee
where $F^{\m\n}$ is a contravariant tensor density
to be varied independently, then insert the usual
3+1 decomposition of the metric into its spatial
part $g_{ij}$, mixed part $g_{0i} \equiv N_i$ and
time-time part $g^{00} \equiv -N^{-2}$, so
that $\sqrt{-g} = N\sqrt{g}$ where $|g|$ is the 
3-metric determinant.  Then it immediately follows
that $I_M$ can be written as \cite{004}
\be
I_M [ {\bf E},{\bf A}]  =  -\int d^4x 
[E^i\dot{A}_i  
+ \textstyle{\frac{1}{2}} \, Ng^{-1/2} 
g_{ij} (E^iE^j +B^iB^j) 
- \e_{ijk}
N^iE^jB^k ]    \label{action1}
\ee
where $F^{0i} \equiv E^i$ is the electric, 
$B^i \equiv \e^{ijk}\partial_j A_k$ the magnetic, 
field (both are contravariant three-densities) and 
all metric operators are in 3-space; 
we have solved the Gauss constraint 
(still $\pa_i E^i =0$) so that both  
$E^i$ and $B^i$ are identically transverse,   
$\partial_i E^i = 0 = \partial_i B^i$.  Note
that although it is on an arbitrary curved 
background space, (14) is easily seen to be 
invariant under (11) via the same (metric 
independent!) generator $G$ of (12) since the canonical 
variables and kinetic term are unchanged while
$({\bf E}^2 + {\bf B}^2)$ and ${\bf E}\times {\bf B}$ 
are clearly locally invariant under (11).

We now turn to the black hole 
case and include electric and magnetic sources. 
To stick to the problem of interest in \cite{003}, 
where only the
exterior solution is considered, one can still work 
with the source-free Maxwell equations but one must 
allow for non-vanishing electric and magnetic fluxes 
at infinity. This is possible because the spatial 
sections $\Sigma$ have a
hole.  There are thus two-surfaces that are
not contractible to a point, namely, the surfaces
surrounding the hole (we assume for simplicity 
a single black hole but the
analysis can straightforwardly be extended to the 
multi-black hole case).

Let us first dispose of a technicality when 
varying in presence of electrical sources or
fluxes. The variation of the action under 
changes of $E^i$, 
\be
\delta_E I_M = -\int d^4x \delta E^i
(\dot{A}_i  + Ng^{-1/2}g_{ij}E^j
- \e_{ijk} B^j N^k),
\ee
vanishes for arbitrary variations $\delta E^i$ 
subject to the transversality 
conditions\footnote{The condition
$\delta \oint _{S^2_{\infty}} E^i dS_i =0$ is actually
a consequence of $\partial_i \delta E^i = 0$ (and
of smoothness) on spatial sections with $R^3$-topology.  
We write it separately, however, because this is no 
longer the case if $\Sigma$ has holes, as below.} 
$\partial_i \delta E^i = 0$
and $\delta \oint _{S^2_{\infty}} E^i dS_i =0$ if and 
only if the coefficient of $\delta E^i$ in (15) 
fulfills the condition 
\begin{equation}
\dot{A}_i  + Ng^{-1/2}g_{ij}E^j- \e_{ijk} B^j N^k=
\partial_i V
\label{equE}
\end{equation}
where $V$ ($\equiv A_0$) is an arbitrary function 
which behaves asymptotically as $C + O(r^{-1})$:  
In that
case, $\delta I_M = -\int d^4x \delta E^i
\partial_i V = - \oint _{S^2_{\infty}} \delta 
E^i V dS_i
= - C \delta$(electric flux) = 0.  No special 
conditions are 
required, on the other hand, when varying $A_i$.  
Thus, (14) is appropriate as
it stands, {\it i.e.}, without ``improving" it by adding
surface terms to the variational principle in which the
competing histories all have the same given electric
flux at infinity and thus also the same given electric
charge (here equal to zero). As pointed out in 
\cite{003},
it is necessary to allow the temporal component $V$ 
of the vector potential to approach a non-vanishing 
constant at infinity since this is what happens in the
black hole case if $V$ is required to be regular on the
horizon.  However, as we have just shown,
in order to achieve this while working with
this action, it is unnecessary to keep all
three components $E^i$ of the electric field fixed at 
spatial infinity;   only  the electric flux
$\oint _{S^2_{\infty}} E^i dS_i$ must be kept constant 
in the variational principle.

In the presence of a non-vanishing magnetic flux, 
the magnetic field is given by the expression 
\begin{equation}
B^i = \e^{ijk} \partial_j A_k + B^i_S
\label{magnetic}
\end{equation}
where $B^i_S$ is a fixed field that carries the 
magnetic flux,
\begin{equation}
\oint _{S^2_{\infty}} B^i_S dS_i = 4 \pi \mu,
\end{equation}
and where $B^i_T = \e^{ijk} \partial_j A_k$ is the 
transverse part of $B^i$,
\begin{equation}
\partial_i B^i_T = 0, \; \oint _{S^2_{\infty}} 
B^i_T dS_i = 0.
\end{equation}
Following Dirac, we can take $B^i_S$ to be entirely
localized on a string running from the source-hole 
to infinity, 
say along the positive $z$-axis $\theta =0$.
We shall not need the explicit form of $B^i_S$ in 
the sequel, 
but only to remember that for a given magnetic 
charge $\mu$,
$B^i_S$ is completely fixed and hence is not a 
field to be varied 
in the action.  The only dynamical components
of the magnetic field $B^i$ are still the transverse 
ones, {\it i.e.}, $A_i$.

One can also decompose the electric field as
\begin{equation}
E^i = E^i_T + E^i_L
\end{equation}
where the longitudinal part carries all the 
electric flux
\begin{equation}
\oint _{S^2_{\infty}} E^i_L dS_i = 4 \pi e
\end{equation}
and the transverse field obeys
\begin{equation}
\partial_i E^i_T = 0, \; \oint _{S^2_{\infty}} 
E^i_T dS_i = 0
\end{equation}
and can thus again be written as $E^i_T = \e^{ijk} 
\partial_j Z_k$ for some $Z_k$.
Given the electric charge $e$, the longitudinal 
electric field is completely determined if we 
impose in addition, say,
that it be spherically symmetric.  As we have done 
above, we shall work with a variational principle 
in which we have solved Gauss's law and in which 
the competing histories have
a fixed electric flux $\oint _{S^2_{\infty}} 
E^i dS_i$ at infinity.  This means that the 
longitudinal electric field is completely frozen 
and that only the tranverse
components $E^i_T$ or $Z^i$ are dynamical, 
as for the magnetic field.

In order to discuss duality, it is convenient to 
treat the non-dynamical components of $E^i$ and
$B^i$ symmetrically.  To that end, one may either 
redefine $B^i_S$ by adding
to it an appropriate transverse part so that it 
shares the spherical symmetry  of $E^i_L$, or one
may redefine $E^i_L$ by adding
to it an appropriate transverse part so that it 
is entirely localized on the string.  Both choices 
(or, actually, any
other intermediate choice) are acceptable here.  
For concreteness 
we may take the first choice; the fields then
have no string-singularity.

In the Maxwell action, 
$E^i$ and $B^i$ are now the {\it total}
electric and 
magnetic fields.  Since $E^i_L$ may be taken to 
be time-independent
(the electric charge is constant), one may replace 
$E^i$ by $E^i_T$ in the kinetic term of (14), 
yielding as alternative action
\be
I_M^{e,\mu} [{\bf E}_T, {\bf A}] = -\int d^4x
[E^i_T\dot{A}_i 
+ \textstyle{\frac{1}{2}} \, Ng^{-1/2}
g_{ij} (E^iE^j +B^iB^j) - \e_{ijk}
N^iE^jB^k ].    \label{action2bis}
\ee
This amounts to dropping a total time derivative -- 
equal to zero for periodic boundary conditions -- 
and shows explicitly that the kinetic term is
purely transverse. Note that there is actually a
{\it different} action (23), hence a distinct 
variational principle, for each choice of $e$ 
and $\mu$, as the notation indicates.

Consider now a duality rotation acting on the 
transverse, dynamical 
variables $A_i$ (or $B^i_T$) and $E^i_T$.  Just as 
in the sourceless 
case, the kinetic term of is invariant
under this transformation: it is the same kinetic 
term and the 
transformation law is the same;
the surface term at the horizon in the variation 
 vanishes because $\dot{A}_i=0$ and 
$\dot{Z}_i=0$ there.\footnote{To discuss the surface 
terms that arise in the variation of the 
action, one must supplement the asymptotic behavior 
of the fields 
at infinity specified earlier by conditions at the 
horizon.  These are especially obvious 
in the Euclidean continuation, where 
time becomes an angular variable with the horizon 
sitting at the 
origin of the corresponding polar coordinate system.  
Regularity then requires that 
$V \equiv A_0$ and the time derivatives 
$\dot{A}_i , \; \dot{E}^i $ all vanish at the horizon.  
We assume these conditions to be fulfilled throughout.}  
Thus, if we also rotate the (non-dynamical) components 
of the electric
and magnetic fields in the same way, that is, if we
relabel the external parameters $e$, $\mu$ by the 
same 2D rotation, so that the 2-vector 
$Q \equiv (e,\m )$, becomes
\begin{equation}
Q^\prime = RQ
\end{equation}
then the actions $I_M^{e,\mu}$ and 
$I_M^{e^\prime,\mu^\prime}$ are equal
since ${\bf E}$ and ${\bf B}$ enter totally 
symmetrically in the energy and momentum densities.
More explicitly, if we write the longitudinal fields 
as $B^i_L = \mu V^i$, $E^i_L = eV^i$, then the 
relevant terms in (23) are just 
\be
- \int d^4x \{ 
Ng^{-1/2} g_{ij} 
[ (e E^i_T  + \mu B^i_T) V^j 
+ \frac{1}{2} (e^2 + \mu^2 ) V^iV^j  ]
- \epsilon_{ijk} N^i V^j (e B^k_T - \mu E^k_T ) 
\} \; .
\ee
For the mixed terms, it is clear that the field 
transformation is just compensated by the parameter 
rotation (24), 
while the $VV$ term is invariant under the latter.  
To put it more formally, 
the extended duality invariance we have spelled 
out is one that links {\it different} systems,
with different parameters:
\begin{equation}
I^{e,\mu}_M [{\bf E}_T, {\bf A}_T] =
I^{e^\prime ,\mu^\prime}_M [{\bf E}_T^\prime, 
{\bf A}_T^\prime]\; ,
\label{invar}
\end{equation}
where the primes denote the rotated values. As a 
special case, for the black holes without 
Maxwell excitations, we find
equality of equally electrically and
magnetic charge actions,
\begin{equation}
I_M^{e,0} [{\bf 0,0}] = I_M^{0,e}[{\bf 0,0}]
\end{equation}
as also obtained, by explicit calculation of these
actions, in \cite{003}.  
This equality is thus not a special artifact, but 
reflects a general
invariance property of the action appropriate to the
variational principle considered here, in which the 
electric and magnetic fluxes are kept fixed.

It is a pleasure to acknowledge my collaborators,
M. Henneaux and C. Teitelboim, as well as
support from the National Science Foundation, under
grant \#PHY-9315811.

\end{document}